# Rotational Damping and Compound Formation in Warm Rotating Nuclei

M. Matsuo[1], S. Leoni[2], C. Grassi[2], E. Vigezzi[2], A. Bracco[2], T. Døssing[3] and B. Herskind[3]

[1] *Graduate School of Science and Technology, Niigata University, Niigata 950-2181, Japan*
[2] *INFN sez. Milan, and Department of Physics, University of Milan, Milan, Italy*
[3] *Niels Bohr Institute, University of Copenhagen, Copenhagen Ø, Denmark*

**Abstract.** The rotational damping width $\Gamma_{rot}$ and the compound damping width $\Gamma_{comp}$ are two fundamental quantities that characterize rapidly rotating compound nuclei having finite thermal excitation energy. A two-component structure in the strength function of consecutive E2 transitions reflects the two widths, and it causes characteristic features in the double and triple gamma-ray spectra. We discuss a new method to extract experimentally values of $\Gamma_{rot}$ and $\Gamma_{comp}$. The first preliminary result of this method is presented.

## INTRODUCTION

Quasicontinuum gamma-rays emitted from compound nuclei formed by heavy-ion fusion reactions carry information on rapidly rotating nuclei which have thermal excitation energy U of an order of 1-8 MeV. The warm nuclei are characterized by a high density of energy levels, which is a consequence of a large number of many-particle many-hole (np-nh) states associated with thermal excitations. In addition, the residual two-body interaction acting among particles and holes leads to formation of compound states. Energy levels of compound states are very different from the levels of shell model character observed in the cold region (U<1MeV); they are complicated mixtures of many np-nh configurations. The compound states at U>8 MeV are known as a typical example of quantum chaos[1,2,3] since their statistical properties of level and strength fluctuations exhibit the generic laws[4] governed by random matrix theories. We thus expect in the region of U=1-8 MeV an interesting order-to-chaos transition, i.e., the transition from simple shell model states to the complicated compound states.

The collective rotation of deformed nuclei exhibits a transition from rotational band structure to rotational damping with increasing thermal excitation energy U[5]. Analysis of quasicontinuum gamma-rays has revealed that the energy levels forming the rotational band structures characterized by strong E2 transitions within band members are limited only in the cold region U<1 MeV, whereas the E2 decay from levels in the warm region U>1 MeV are spread over many decay branches, implying the damping of collective rotation[6,7]. The onset of rotational damping is a consequence of the formation of compound states and it is closely related to the order-to-chaos transition expected in the same energy region. We thus expect that the quasi-continuum E2 gamma-rays can be used as a probe to study the formation of compound states in rotating nuclei. We report here our recent theoretical and experimental efforts to this end.

## ROTATIONAL AND COMPOUND DAMPING WIDTHS

The rotational damping width $\Gamma_{rot}$ and the compound damping width $\Gamma_{comp}$ are two fundamental quantities. $\Gamma_{rot}$ measures the spreading of the E2 strength associated with gamma-decays from an energy level exhibiting rotational damping[5]. It is defined as the FWHM of the B(E2) strength distribution measured by the gamma-ray energy (Fig.1 left). This quantity is related to a time scale of decoherence in the damped collective rotation. The compound damping width $\Gamma_{comp}$, on the other hand, is a measure of the compound state formation. Taking a typical np-nh state at spin I, we consider how the strength of this state is distributed among the compound energy levels at the same spin I. The energy interval of this strength distribution is $\Gamma_{comp}$ (Fig.1 right). In other words, $\Gamma_{comp}$ is the spreading width of np-nh states, which is caused by configuration mixing due to the residual interaction. Its reciprocal is related to a relaxation time for np-nh states to form compound states. $\Gamma_{comp}$ is expected to increase with thermal excitation energy U.

The two widths $\Gamma_{rot}$ and $\Gamma_{comp}$ play central roles in the dynamics of warm rotating nuclei[5]. Rotational band structure is realized if $\Gamma_{comp}, \Gamma_{rot} < D$, where D is the mean level spacing (the notion of damping becomes meaningless in this case). The opposite condition $\Gamma_{comp}, \Gamma_{rot} > D$ characterizes the onset of configuration mixing and the onset of rotational damping. If $\Gamma_{comp} > \Gamma_{rot} > D$, the strong mixing brings about the *motional narrowing effect*[5], which gives counter-intuitive decrease of $\Gamma_{rot}$ as a function of U. In an extreme case of $\Gamma_{comp} \gg D > \Gamma_{rot}$, there may emerge the *ergodic rotational bands*, where the strongly chaotic compound energy levels form rotational bands instead of exhibiting rotational damping[8]. It is therefore important to evaluate $\Gamma_{rot}$ and $\Gamma_{comp}$ to understand the warm rotation.

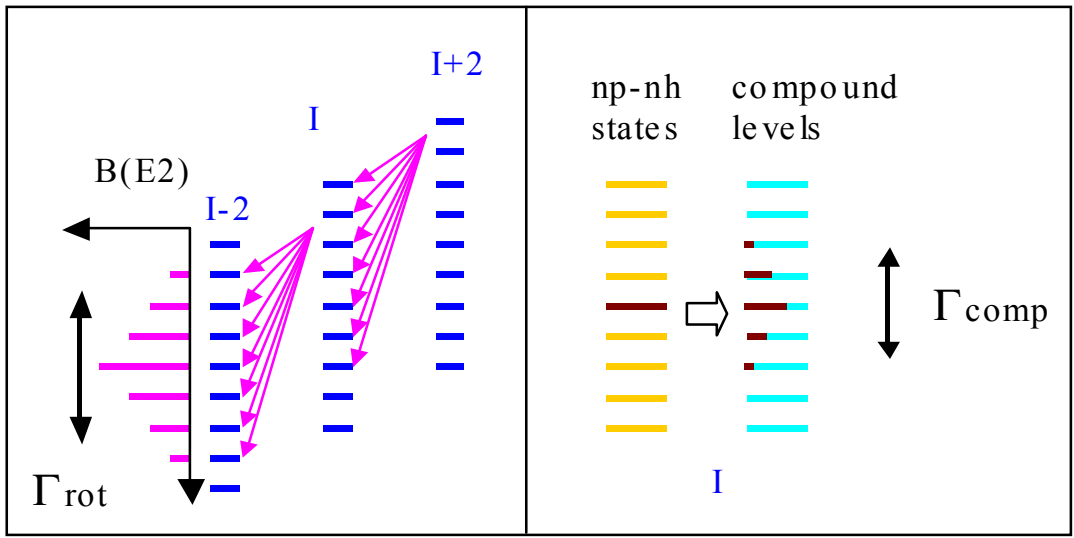

**FIGURE 1.** The rotational damping width $\Gamma_{rot}$ (left) and the compound damping width $\Gamma_{comp}$ (right).

Theoretical predictions of $\Gamma_{rot}$ and $\Gamma_{comp}$ can be provided by the microscopic cranked shell model calculations[9,10]. In this model the intrinsic excitation is described on the basis of particle-hole excitations in a uniformly rotating deformed mean-field whereas the rotational angular momentum is treated approximately. Each np-nh configuration is ascribed to an unperturbed rotational band. As the thermal excitation energy U increases, configuration mixing among np-nh unperturbed bands becomes strong. Using the shell model diagonalization technique, we obtain energy levels and their wave functions for U< a few MeV with a reasonable choice of cut-off energy of basis truncation. Near the yrast line, the energy levels after diagonalization still retain a rather pure cranked shell model character and form rotational bands. The rotational damping sets in around U> 1 MeV, leaving only 20-30 rotational bands in the case of a typical deformed nuclei $^{168}$Yb[10], in agreement with the experimental observation[6,7].

$\Gamma_{rot}$ is governed by rotational alignment of the single-particle orbits and by the compound damping width $\Gamma_{comp}$. If the mixing is not very strong ($\Gamma_{comp} < \Gamma_{rot}$), the rotational damping width $\Gamma_{rot}$ is dominated by the single-particle alignments[5]. The alignments exhibit a strong shell effect which varies depending on the rotational frequency, the deformation, and the position of the Fermi level (i.e. the neutron and proton numbers) on top of a smooth dependence on these quantities[11]. The cranked shell model calculation predicts $\Gamma_{rot}$ =100-300 keV (increasing with spin) in a typical rare-earth nucleus $^{168}$Yb, but it may vary within the very wide interval $\Gamma_{rot}$ =30-500 keV, depending on nuclear species[12]. An extremely small value $\Gamma_{rot}$ ~30keV is calculated for the superdeformed $^{192}$Hg, where the shell effect is very strong[11].

Experimentally the value of $\Gamma_{rot}$ has been difficult to measure. A previous analysis [13] using the technique of the rotational plane mapping of the 3D spectra of triple coincident gamma-rays has suggested an order of 50-80 keV in $^{168}$Yb, but at the same time it was pointed out that the spectra are not consistently described by the simple assumption of a single Gaussian or Lorentzian profile of E2 strength functions, that is adopted in the analysis. On the other hand, a recent analysis[14] using a different technique reports $\Gamma_{rot}$ =200-300 keV in $^{168}$Yb, in good agreement with the theory.

Concerning $\Gamma_{comp}$, the model predicts that it is rather independent on nuclear species, deformations and spins. A typical value is $\Gamma_{comp} \sim$ 50 keV in the energy region U~1-2 MeV[12,15]. Experimentally, however, there has been no direct measurement so far, not only in rotating nuclei but also in connection with other phenomena. This difficulty arises from a fact that there is no direct way to excite shell model np-nh states. Nevertheless, we have recently shown [15] that it should be possible to measure both $\Gamma_{rot}$ and $\Gamma_{comp}$ by a properly designed analysis. Indeed in the present work we extract for the first time experimental value, though preliminary, of the compound damping width $\Gamma_{comp}$

## NARROW COMPONENT IN 2D AND 3D SPECTRA

To probe $\Gamma_{rot}$ and $\Gamma_{comp}$, we utilize a theoretical observation[12,15] that the E2 strength function reflects both $\Gamma_{rot}$ and $\Gamma_{comp}$ through a doorway mechanism of rotational damping (although this is only valid when $\Gamma_{rot} > \Gamma_{comp}$). In an illustration in Fig.2(a), we suppose that an initial level at spin I is a mixture of three unperturbed np-nh bands. The damped E2 transition from this level follows the unperturbed transition energies of these unperturbed bands. The unperturbed np-nh bands, however, are not energy eigenstates, and mix with other np-nh states at spin I-2, leading to smearing by the spreading width $\Gamma_{comp}$. As a consequence, the E2 strength function exhibits fine structures which are characterized by $\Gamma_{comp}$ whereas the overall profile is determined by $\Gamma_{rot}$. Considering further two-dimensional (2D) distribution of E2 strength for two consecutive E2 decays γ1 and γ2 passing I → I-2 → I-4, the fine structure brings a characteristic correlation along the diagonal line in the $E_{γ1} \times E_{γ2}$ plane (Fig.2(b)). It is a remnant of the rotational correlation in the unperturbed np-nh bands, but the correlation is smeared by $\Gamma_{comp}$. Cutting the 2D strength distribution in the direction perpendicular to the diagonal, the strength plotted in the $E_{γ1}$- $E_{γ2}$ axis exhibits a two-component structure centered around the ridge position $E_{γ1}$- $E_{γ2}$=4/$I$≈70 keV($I$ being the moment of inertia) with a narrow distribution with width $\Gamma_{narrow}$, and a wide distribution with width $\Gamma_{wide}$ (Fig.2(c)). The narrow component arises from the doorway mechanism. One can relate the width $\Gamma_{narrow}$ and the intensity $I_{narrow}$ of the narrow component to the compound damping width by $\Gamma_{narrow} \sim 2\Gamma_{comp}$, and $I_{narrow} \sim 2D/\Gamma_{comp}$. The width $\Gamma_{wide}$ of the wide component can be related to the rotational damping width $\Gamma_{wide} \sim \sqrt{2}\Gamma_{rot}$ assuming a Gaussian shape for the overall E2 distribution ($\Gamma_{wide} \sim 2 \Gamma_{rot}$ if we assume a Lorentzian shape). See also [13].

Actual spectra contain not only the consecutive gamma-rays but also non-consecutive ones which form distributions around the second, third etc. ridge positions $E_{γ1}$- $E_{γ2}$=70×n keV (n=2,3, …). Symmetric combinations exchanging γ1 and γ2 should be included also. Fig.3(a) shows a simulated spectrum calculated with use of the cranked shell model and a simple simulation of cascades neglecting the E1 contribution. Here a sizable structure observed around the first ridge $E_{γ1}$-$E_{γ2}$=±70keV originates from the narrow component.

The narrow component creates more prominent features in the 3D spectra of coincident three gamma-rays, as shown in Fig.3(d). The figure is a 2D representation of the spectra projected on the 'transversal plane', defined by $E_{γ1}+E_{γ2}+E_{γ3}$=const, which is perpendicular to the diagonal line in the $E_{γ1} \times E_{γ2} \times E_{γ3}$ cube (Fig.3(c)). A volcano-like structure characterized by a crater rim and six peaks around, and ridges of volcano extending from the peaks is seen. The rim-ridge structure arises from the narrow component.

The displaced rotational plane cut, which we have developed recently[16,17], is a method to map the volcano-like structure of the 3D spectra. The displaced rotational plane, defined by $E_{γ1}+E_{γ2}-2E_{γ3}$=δ, is perpendicular to the transversal plane (Fig.3(f)). The plane without a displacement (i.e. δ=0) intersects the volcano at the center. By displacing the plane with variation of δ, the volcano-like structure can be scanned. See [16,17] for results of this method.

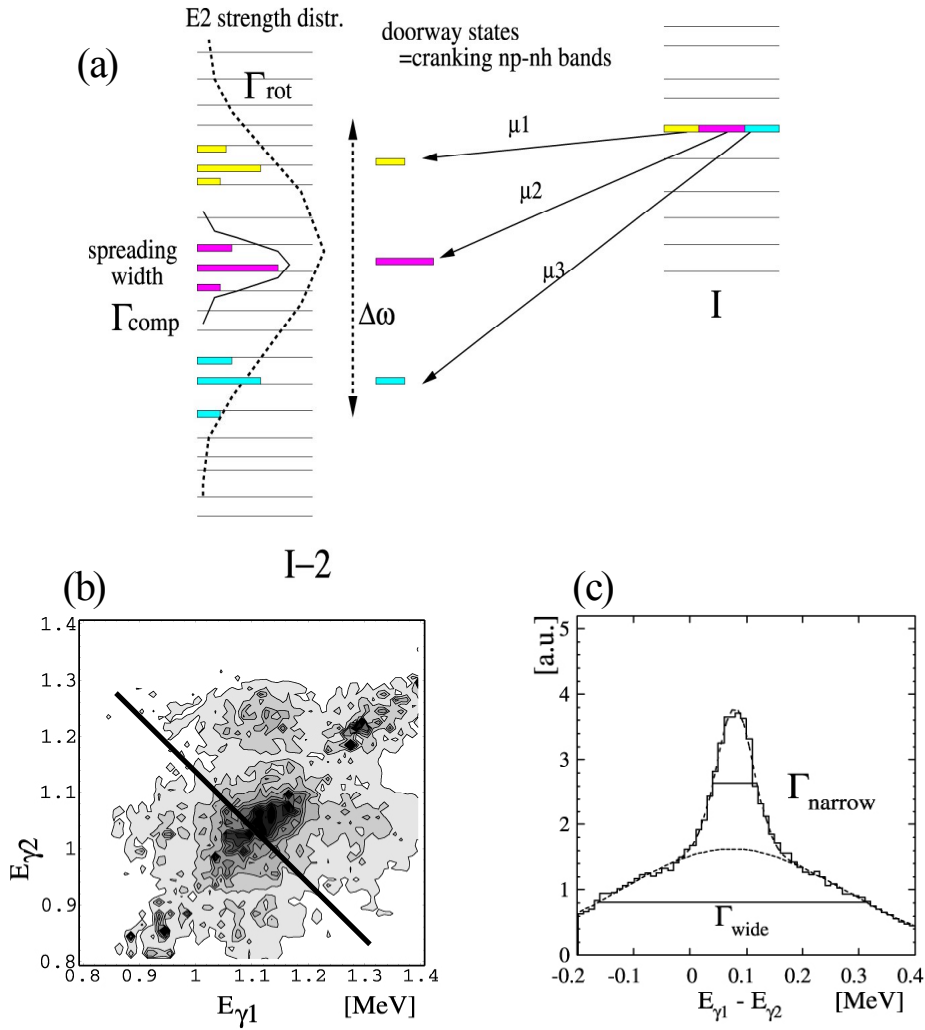

**FIGURE 2.** (a) A schematic illustration of the doorway mechanism of the rotational damping, as explained in the text. (b) An example of the strength function for consecutive two E2 gamma-rays, calculated microscopically by means of the cranked shell model. Here the strength function is averaged over levels in an energy bin including 51-st to 100-th levels (for each $I^\pi$) at spin I=42,43 in $^{168}$Yb. The thick bar is a perpendicular cut. (c) The strength projected on the perpendicular cut.

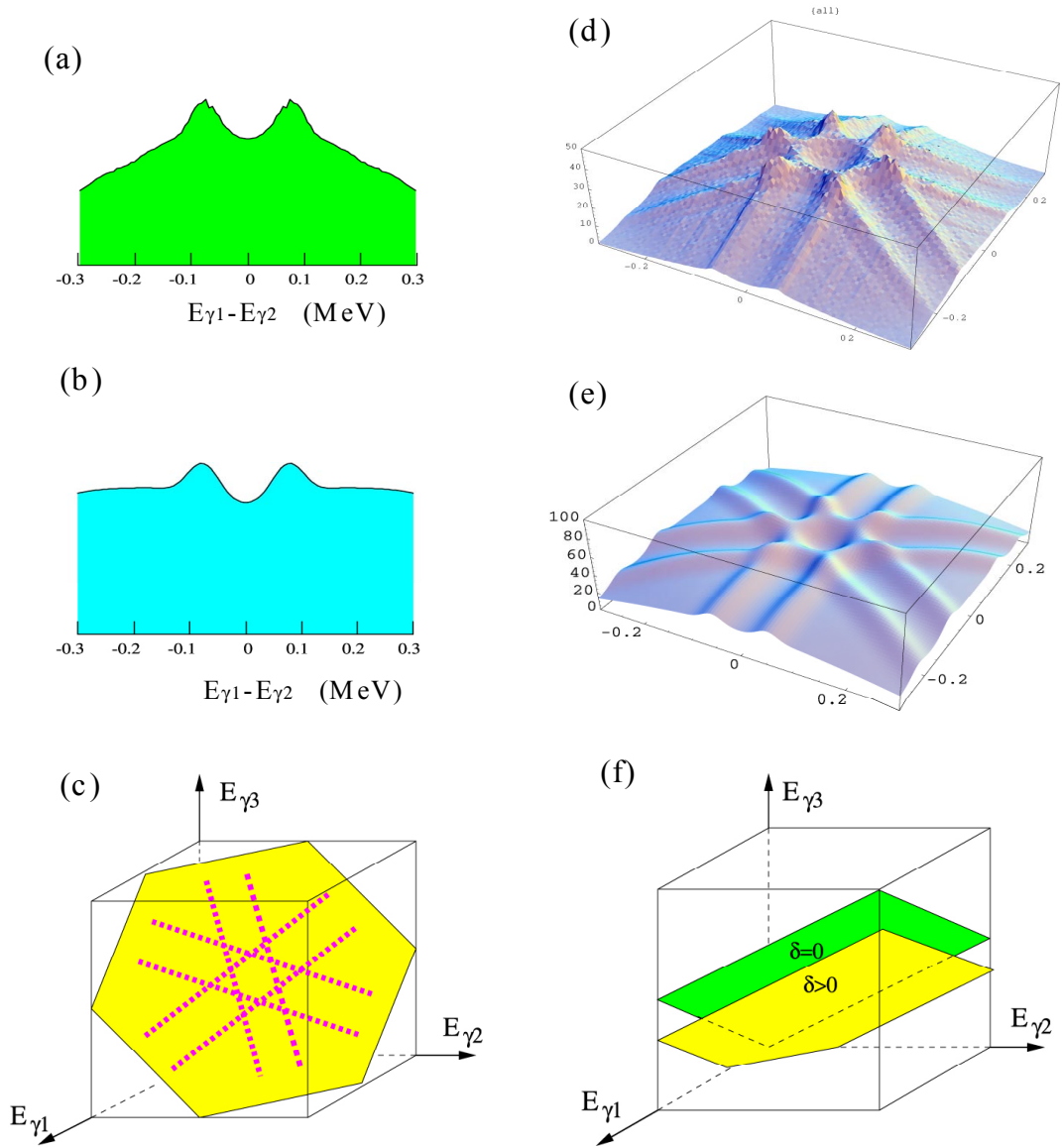

**FIGURE 3.** (a) The 2D spectra calculated microscopically with use of the cranked shell model (see text), projected on the perpendicular cut, and plotted as a function of Eγ1-Eγ2. Included are only cascades passing through the excited levels (the 51 to 100 levels at each $I^\pi$) in spin interval I=34-43 in $^{168}$Yb. (b) A parameterized 2D spectrum which incorporates the narrow component (see text). The used parameter values Γrot =287keV, Γnarrow=67keV, and Inarrow=20% are taken from a fit to the strength function of consecutive E2 transitions (see Fig.2(c) also) for the same energy levels as in (a). (c) A transversal plane in the 3D spectra of triple coincident gamma-rays. (d) The 3D spectra projected on the transversal plane, calculated with use of the cranked shell model for the same energy levels as (a). (e) A parameterized 3D spectrum in the transversal plane. The parameter values are the same as (b). (f) Displaced rotational planes.

# EXTRACTING TWO DAMPING WIDTHS

Motivated by these studies, we recently have developed another new analysis aiming at extracting the rotational and compound damping widths from experimental quasi-continuum gamma-ray spectra. For this purpose we introduce a parameterized spectrum function which incorporates the two-component structure. The strength function for the consecutive two E2 gamma rays is given as

$$S(E_{\gamma 1}, E_{\gamma 2}) = I_{\text{narrow}} S^{\text{correl}}(E_{\gamma 1}, E_{\gamma 2}) + (1 - I_{\text{narrow}}) S^I(E_{\gamma 1}) S^{I-2}(E_{\gamma 2}). \quad (1)$$

Here $S^I(E_{\gamma 1})$ (and $S^{I-2}(E_{\gamma 2})$) is a strength function for single step decay from states at spin I (and I-2), which is a simple Gaussian function whose FWHM is set to the rotational damping width $\Gamma_{\text{rot}}$. The second term represents the uncorrelated part in the consecutive transitions and corresponds to the wide component. The first term on the other hand represents the narrow component which is correlated along the diagonal line and has a narrow width to the $E_{\gamma 1}$-$E_{\gamma 2}$ direction (cf. Fig.2(b)). We parameterize $S^{\text{correl}}(E_{\gamma 1}, E_{\gamma 2})$ by a two-dimensional Gaussian function which contains two parameters; the width $\Gamma_{\text{narrow}}$ of the narrow component and the rotational damping width $\Gamma_{\text{rot}}$. A projection of $S(E_{\gamma 1}, E_{\gamma 2})$ on the perpendicular cut exhibits two-component Gaussian distribution similar to Fig.2(c) whereas projections on the $E_{\gamma 1}$ and $E_{\gamma 2}$ axes have the same distributions as $S^I(E_{\gamma 1})$ and $S^{I-2}(E_{\gamma 2})$, respectively.

For multiple steps of E2 gamma decays, we consider a multidimensional strength function $S(E_{\gamma 1}, E_{\gamma 2}, \cdots, E_{\gamma n})$ given by a conditional product of the two-step functions (1). Spectra for double and triple coincident gamma-rays are obtained by integrating unobserved gamma-rays. These procedures can be performed analytically, leading to final functional forms of 2D and 3D spectra which consist of many Gaussian functions. The spectra thus obtained contain four parameters; two damping widths $\Gamma_{\text{rot}}$ and $\Gamma_{\text{narrow}}$ (related to $\Gamma_{\text{comp}}$), the intensity $I_{\text{narrow}}$ of the narrow component, and the number of decay steps considered. We confirm that the parameterized spectra describe very well the essential features of the cranked shell model spectra, as seen by comparing Fig.3(a) and (b), and (d) and (e).

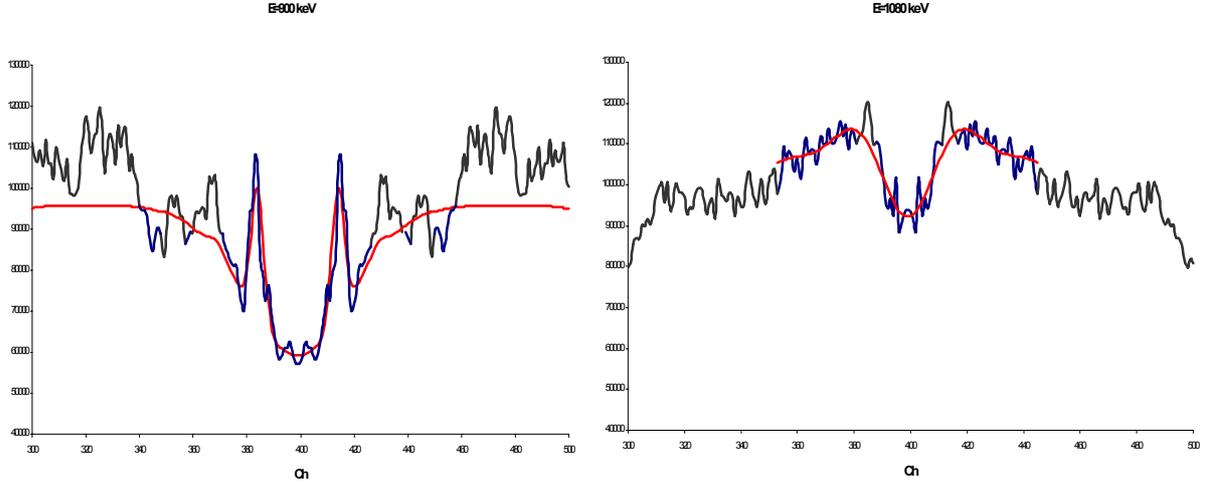

**FIGURE 4.** Experimental 2D spectra in the perpendicular cut at E$\gamma$=900 and 1080 keV in $^{168}$Yb. The parameterized spectra fitted to the data are also shown.

By using a multidimensional fitting of the parameterized spectrum to the experimental E2 spectrum, we are able to extract the best fit values of $\Gamma_{\text{rot}}$, $\Gamma_{\text{narrow}}$, and $I_{\text{narrow}}$. As a first attempt, we applied this method to the perpendicular cut of the 2D spectrum in $^{168}$Yb[18] at different gamma-ray energies E$\gamma$=(E$\gamma$1+E$\gamma$2)/2. The known discrete peaks are subtracted from the data. The E1 contribution to quasi-continuum spectra is evaluated with use of a spectrum function $\propto E_\gamma^3 \exp(-E_\gamma / T)$, and subtracted. In the fitting the peak regions close to the ridge positions are neglected in order to remove contribution from unresolved discrete rotational bands, but this does not affect very much the extracted values of parameters. Examples of the experimental spectra and the parameterized spectra fitted are shown in Fig.4. The extracted parameter values thus determined are listed in Table1.

**TABLE 1. Extracted parameter values in $^{168}$Yb**

| E$\gamma$ [keV] | 900 | 960 | 1080 | 1140 |
|---|---|---|---|---|
| $\Gamma_{rot}$ [keV] | 125 | 127 | 171 | 145 |
| $\Gamma_{narrow}$ [keV] | 22 | 38 | 96 | 53 |
| $I_{narrow}$ [%] | 13 | 22 | 38 | 9 |

This preliminary result is compared with the cranked shell model calculation in Fig. 5, where the recent report by the Berkeley group [14] is also included. In the model calculation, the rotational damping width $\Gamma_{rot}$ and the width of narrow component $\Gamma_{narrow}$ are evaluated from the E$\gamma$1-E$\gamma$2 correlation (cf. Fig.2(c)) calculated for the lowest 300 energy levels for each spin and parity, covering approximately the region up to U~2.3MeV. ($\Gamma_{rot}$ is obtained with use of $\Gamma_{wide}$ and the relation $\Gamma_{rot}=\Gamma_{wide}/\sqrt{2}$ ). In this figure, the gamma-ray energy is converted to the rotational spin by using the experimental average gamma-ray energy[14].

The extracted values of the rotational damping width $\Gamma_{rot}$ agree fairly well with the model. In a previous analysis[13], we estimated $\Gamma_{rot}$ =50-80 keV as a lower limit in the same region of gamma-ray energy. The present values are much larger than the previous ones. It should be noted that in the previous analysis we have used a different function which neglects the presence of the narrow component.

The experimentally extracted value of $\Gamma_{narrow}$, which we obtain for the first time, is within the range of 20-100 keV, in qualitative agreement with the model prediction (Fig.5 right). Possible contribution from unresolved discrete rotational bands could influence results. An analysis including 3D spectra is in progress. The aim is to subject the two-component parameterization to more stringent tests and to obtain more firm results.

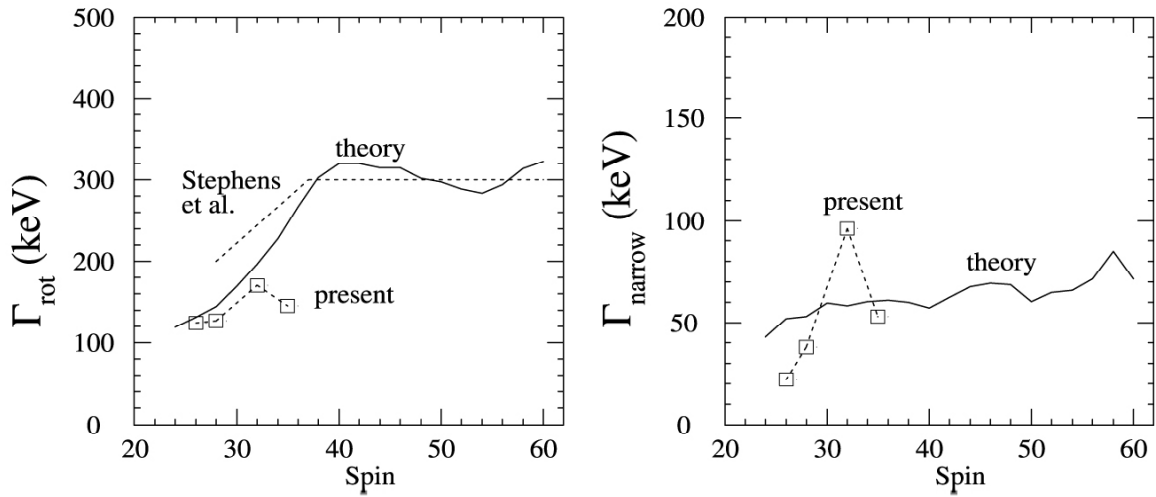

**FIGURE 5.** Comparison of the extracted and theoretical values of $\Gamma_{rot}$ (left) and $\Gamma_{narrow}$ (right). The solid curve is the result of the cranked shell model. The squares represent those obtained in the present data analysis using the parameterized spectra. The result by the Berkeley group[14] is also shown with dotted line for $\Gamma_{rot}$.

## CONCLUSIONS

Compound nuclei with high rotational angular momentum and with finite thermal excitation energy are studied both theoretically and experimentally by analyzing the quasicontinuum E2 gamma-rays spectra. There exits two fundamental quantities; the rotational damping width $\Gamma_{rot}$ which measures the energy spreading in rotational E2 transitions, and the compound damping width $\Gamma_{comp}$ which is the spreading width of many-particle many-hole states carrying thermal excitation. The strength function of consecutive E2 transitions exhibits a two-component structure made of wide and narrow distributions. The width of the narrow component is directly related to $\Gamma_{comp}$ whereas the wide component reflects $\Gamma_{rot}$. The presence of narrow component causes characteristic features in the 2D and 3D spectra of double and triple coincident gamma-rays. We introduced a simple function that parameterizes these features. By fitting the new function to the spectra, it is possible to extract $\Gamma_{rot}$ and $\Gamma_{comp}$ experimentally. We made the first preliminary application of this technique, and derived experimental values of $\Gamma_{comp}$ for the first time.